\documentclass[11pt,reqno,a4,onecolumn,twoside,final,spanish]{article}

\usepackage{amsthm,amsmath,amssymb,latexsym,amsfonts,amscd} 
\usepackage{latexsym}
\usepackage{url}
\usepackage[T1]{fontenc}
\usepackage{ae}
\usepackage{aecompl}
\usepackage[small,sc,hang,bf]{caption} 
\usepackage{graphicx,color,ae,fancyvrb}
\usepackage{pst-all}
\usepackage{subfig}
\usepackage{float} 
\usepackage{verbatim}
\usepackage{enumerate}
\usepackage{array}
\usepackage{longtable}
\usepackage{multicol}
\usepackage{cancel}
\newcommand*\chancery{\fontfamily{pzc}\selectfont}

\newlength\mytemplen
\newsavebox\mytempbox

\makeatletter
\newcommand\mybluebox{%
    \@ifnextchar[
       {\@mybluebox}%
       {\@mybluebox[0pt]}}

\def\@mybluebox[#1]{%
    \@ifnextchar[
       {\@@mybluebox[#1]}%
       {\@@mybluebox[#1][0pt]}}

\def\@@mybluebox[#1][#2]#3{
    \sbox\mytempbox{#3}%
    \mytemplen\ht\mytempbox
    \advance\mytemplen #1\relax
    \ht\mytempbox\mytemplen
    \mytemplen\dp\mytempbox
    \advance\mytemplen #2\relax
    \dp\mytempbox\mytemplen
    \colorbox{myblue}{\hspace{1em}\usebox{\mytempbox}\hspace{1em}}}

\makeatother

\usepackage{natbib}

\usepackage{hyperref}

\usepackage{geometry}
\usepackage{calc}
	
\makeatletter
\renewcommand{\section}{\@startsection{section}{1}{0pt}{-3ex plus -1ex minus 0ex}{2ex plus 0ex}{\bf}}
\makeatother

\makeatletter
\renewcommand{\subsection}{\@startsection{subsection}{1}{0pt}{-2ex plus -1ex minus 0ex}{2ex plus 0ex}{\bf}}
\makeatother

\theoremstyle{definition}

\theoremstyle{remark}

\begin{document}

\renewcommand{\tablename}{Tabla}
\noindent

\begin{flushleft}
\textsl {\chancery  Memorias de la Primera Escuela de Astroestad\'istica: M\'etodos Bayesianos en Cosmolog\'ia}\\
\vspace{-0.1cm}{\chancery  9 al 13 Junio de 2014.  Bogot\'a D.C., Colombia }\\
\textsl {\scriptsize Editor: H\'ector J. Hort\'ua}\\
\href{https://www.dropbox.com/sh/nh0nbydi0lp81ha/AACJNr09cXSEFGPeFK4M3v9Pa}{\tiny {\blue Material suplementario}}
\end{flushleft}



\thispagestyle{plain}\def\@roman#1{\romannumeral #1}



\begin{center}\Large\bfseries Helical  magnetic fields via  baryon asymmetry \end{center}
\begin{center}\normalsize\bfseries  Campos magn\'eticos con helicidad v\'ia asimetr\'ia bari\'onica.\end{center}

\begin{center}
\small
\textsc{Eduard F. Piratova \footnotemark[1], Edilson A. Reyes \footnotemark[1] \& H\'ector J.  Hort\'ua \footnotemark[1]}
\footnotetext[1]{Departamento de Ciencias B\'asicas, Fundaci\'on Universitaria Los Libertadores}

\end{center}
\noindent\\[1mm]
{\small
\centerline{\bfseries Abstract}\\

There is strong observational evidence for the presence of large-scale magnetic fields MF in galaxies and clusters, with strength
 $\sim\mu G$ and coherence lenght on  the order of Kpc. However its origin remains as an outstanding problem. 
One of the possible explanations is that they have been generated in the early universe.  Recently, it has been proposed  that  helical primordial magnetic fields PMFs,  
could be generated  during the EW or QCD phase transitions, parity-violating processes and predicted by GUT or string theory. 
Here we concentrate on the study of two  mechanisms to generate  PMFs, the first one is the  $\nu$MSM which  triggers instability in the
Maxwell's equations and leads to the generation of  helical PMFs. The second one is the usual electroweak baryogenesis scenario. 
Finally, we  calculate the exact power spectra of these helical PMFs and we show  its role in the production of gravitational waves 
finding a scale-invariant  on large scales  and  an oscillatory motion (damping) for $k\eta\gg 1$. \\

{\footnotesize
}

\noindent\\[1mm]
{\small
\centerline{\bfseries Resumen}\\
Existe una fuerte evidencia observacional de la presencia de campos magn\'eticos en grandes escalas, tanto en galaxias y c\'umulos 
gal\'acticos con intensidades de $\sim\mu G$  y con longitudes de coherencia de Kpc.
Sin embargo, el origen de estos campos magn\'eticos es todavia un problema abierto. 
Una de las explicaciones es que este campo fue generado en etapas tempranas del universo.
Recientemente, se ha propuesto  campos magn\'eticos con helicidad, que pudieron ser generados en transiciones de  
fase electod\'ebil o de QCD,  procesos de violaci\'on de paridad y en  teor\'ias de gran unificaci\'on y teor\'ia de cuerdas.
En este trabajo, se estudia  dos mecanismos que generan campo magn\'etico con helicidad. El primero
es el modelo $\nu$MSM activa la inestabilidad en las ecuaciones de Maxwell y conduce a la generaci\'on 
de campos magn\'eticos con helicidad y el segundo mecanismo bariogenesis electod\'ebil. 
Por \'ultimo, se calcula de una forma exacta el espectro de potencias estos campos con helicidad y se muestra la
producci\'on de ondas gravitacionales encontrando algunas propiedades en escalas sub y super horizonte.  \\}

{\footnotesize
\textbf{Palabras clave:}
Campos magn\'eticos primordiales, asimetr\'ia bari\'onica del universo.\\}

\section{Electroweak baryogenesis}
Before the decoupling epoch the electric conductivity of the plasma $\sigma$ is very large, therefore we can consider the infinite conductivity limit. In this
limit, the induced electric field is zero and the  evolution of a primordial magnetic fields is given by $B(x, \eta) \sim B(x)/a^2(\eta)$,  besides, the magnetic flux and the magnetic helicity are
conserved, 
\begin{equation}
\frac{d}{d\eta}\int \pmb{B}\cdot d\pmb{S}=-\frac{1}{\sigma}\int\nabla \times \nabla\times\pmb{B}\cdot d\pmb{S},
\end{equation}
\begin{equation}
\frac{d }{d\eta}\mathcal{H}_M=-\frac{1}{\sigma}\int\pmb{A}\cdot \pmb{B}dV,
\end{equation}
where $A$ is the vector potential and $\mathcal{H}_M$ is the magnetic helicity which is gauge invariant
\begin{equation}
\mathcal{H}_M=\frac{1}{V}\int \pmb{A}\cdot \pmb{B}dV.
\end{equation}
The magnetic helicity is a very important quantity in astrophysics for different reasons,  \cite{graso}. First, $\mathcal{H}_M$ coincides with the Chern-Simon number $N_{CS}\equiv \int\pmb{A}\cdot \pmb{B}dV $ which is  related to the topological properties of the gauge fields, indeed, a  Chern-Simons term different from zero in Maxwell's  equations  leads to an instability and generation of magnetic fields. The second one, is based on the property  from MHD where the presence of  helicity can transfer magnetic field power (more efficient) to larger length scales, \textit{inverse cascade}. Therefore  we could explain the large magnetic field observed so far. Differents authors have studied generation of PMF via baryon asymmetry, for example in the electroweak context, the violation of the baryon number ($n_L-n_R$) is caused by  sphalerons decay  and  the amoung of helical PMF created can be estimated from
\begin{equation}\label{eq.4}
\frac{d(n_L-n_R)}{dt}\sim -\alpha\frac{d\mathcal{H}_M}{dt}
\end{equation}
where $ \alpha= 1/137$ is the fine structure constant. Other alternative for generating PMF relies in the interaction of the hypercharge component of the electromagnetic field with the axion by means of the anomaly \cite{graso}. The  Lagrangian for this model  can be written as 

\begin{equation}
 \mathcal{L}\sim -\frac{1}{2}\partial_\mu\theta\partial^\mu\theta-\frac{1}{4}F_{\mu\nu}F^{\mu\nu}+\alpha \theta F_{\mu\nu}\tilde{F}^{\mu\nu},
\end{equation}
 where $\theta=\phi_a/f_a$, $\phi_a$ is the axion field, $f_a$ is the Peccei-Quinn symmetry breaking scale and $F_{\mu\nu}$ is the electromagnetic field.  \cite{graso} shows that in  presence of QCD sphaleron the axion equation of motion is 
\begin{equation}
\ddot{\phi}+(3H+\frac{\alpha^4T^3}{f_a^2})\dot{\phi}=0.
\end{equation}
 Using  the later equation,  \cite{graso} estimated that in the tempera\-ture range $1 \text{GeV} > T > 10$MeV and  $f_a > 109$ GeV, the existence of a small PMF  is  possible (with the  QCD sphaleron  out of thermal equilibrium) of the order of
$B\cdot A\sim10^{-22}$. 

\section{The $\nu$MSM}
The Standard Model answers many of the questions about the structure and stability of matter, but it's not complete,  there are still many unanswered questions like neutrino oscillation, matter-antimatter assymetry and dark matter among others, \cite{12}.  The model  $\nu$MSM  was  introduced for overcome some of this problems. In this model, the tiny values of the neutrino masses are related to the small Yukawa Coupling cons\-tants between sterile neutrinos and left-handed leptonic doublets.
 In this model, three singlets ($N_R$) are introduced, these mix with the standard neutrinos and we get the following mass Lagrangian

\begin{equation}
-\mathcal{L}_{\text{mass}}=\frac{1}{2}
\begin{pmatrix}
\overline{(v_L)^c}&\overline{N_R}\\
\end{pmatrix}
\begin{pmatrix}
0&m_D^T\\
m_D&M_R\\
\end{pmatrix}
\begin{pmatrix}
v_L\\
(N_R)^c\\
\end{pmatrix}
+\text{h.c.},
\end{equation}

where $v_L$ are the standard neutrinos, $m_D$ are the dirac-mass matrix and $M_R$ are the Majorana-mass matrix. When this matrix is diagonalized, we get the following physics states

\begin{equation}
\begin{pmatrix}
\xi_1\\
\xi_2\\
\end{pmatrix}
=
\begin{pmatrix}
V_{PMNS}^\dag v-V_{PMNS}^\dag B_1N^c\\
B_1^\dag v+N^c \\
\end{pmatrix}_L \\
+
\begin{pmatrix}
V_{PMNS}^T v^c-V_{PMNS}^T B_1^*N\\
B_1^T v^c+N\\
\end{pmatrix}_R,
\end{equation}

where $V_{PMNS}$ is the Pontecorvo-Maki-Nakagawa-Sakata matrix and  $B_1^\dag=M_R^{-1}m_D $ is a seesaw-factor. $\xi_1$ are the light neutrinos corresponding to standard model neutrinos, while $\xi_2$ are the heavy neutrinos. One of these heavy neutrinos ($N_1$) will be chosen like dark matter candidate while the other two ones ($N_2$, $N_3$) will generate the oscillation neutrinos patterns and will generate baryon asymmetry. This choice generates conditions between the values of its mass and  mixing angles. In particular, if we consider the masses of the light neutrinos  on the order of  mass particles from standard model, the interaction should be \textit{superweak}. The constraint of these free parameters, which comes from cosmology, astrophysics and neutrino oscillation, are shown  in the figure \ref{fig2}. 
\begin{figure}
\centering
\includegraphics[scale=0.9]{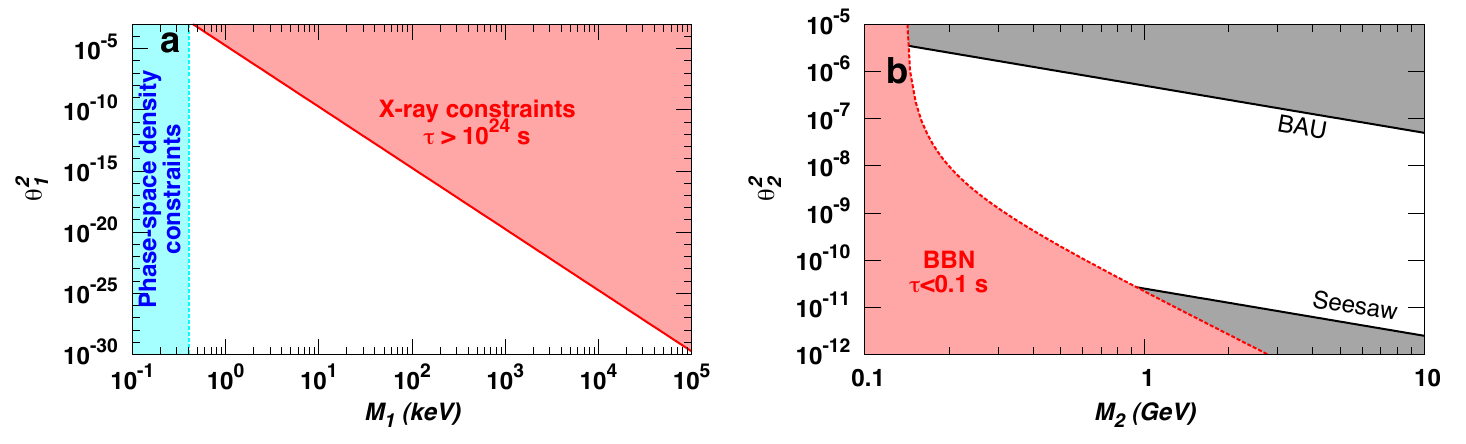}
\caption{Constrains of the neutrinos mass and mixing angles for $N_1$ (left) and $N_{2,3}$ (right), \cite{guia1}.}\label{fig2}
\end{figure}

An important characteristic of this model, is that baryon asymmetry could be possible via leptogenesis. Indeed,  if the three Sakharov conditions are satisfied \cite{sakharov}, the initial lepton asymmetry  generated  by  violating CP in neutrino oscillation, is converted to baryon asymmetry through electroweak anomaly (with  field configurations  no-conserved baryon number -\textit{sphalerons}-).  In this way, the helical PMF is produced as we see in  equation (\ref{eq.4}).
\paragraph{The  statistics for  helical PMFs\\}
We consider a stochastic PMF  generated before  recombination. The power spectrum  which is defined as the Fourier transform of the two point correlation  can be written as
\begin{equation} 
\langle B_j({\mathbf k})B^{*}_l({\mathbf k'})\rangle 
=(2\pi)^3 \delta({\mathbf k}-{\mathbf k'}) [P_{jl} S(k)   + 
i \epsilon_{jlm} \hat{k}_m A(k)], 
\label{spectrum} 
\end{equation} 
here $S (k)$ and $A (k)$  are  the symmetric and helical part of the PMF power spectrum respectively, $P_{ij} = \delta_{ij} - \hat{k}_i\hat{k}_j$ is the  transverse plane projector and 
 $\epsilon_{ijl}$ is the antisymmetric tensor. We scale the PMF  by a power law (for $k<k_D$)
\begin{equation}
S(k) = \frac{B^{2}_\lambda (2\pi)^2\lambda^{n_S+3}}{\Gamma\left(\frac{n_S+3}{2}\right)} ~k^{n_S}, \quad A(k) = \frac { B^2_\lambda(2\pi)^2\lambda^{n_A+3}}{\Gamma\left(\frac{n_A+4}{2}\right)} ~k^{n_A}, 
\end{equation}
where   $n_S$, $n_A$, are the spectral indices for  symmetric and helical parts
respectively and  $k_{D}$ is an ultraviolet cut-off  (for a  dependence of an infrared cutoff and its dependence with spectral index, see \cite{hector}).
\begin{figure}[h!]
  \centering
    \includegraphics[width=.9\textwidth]{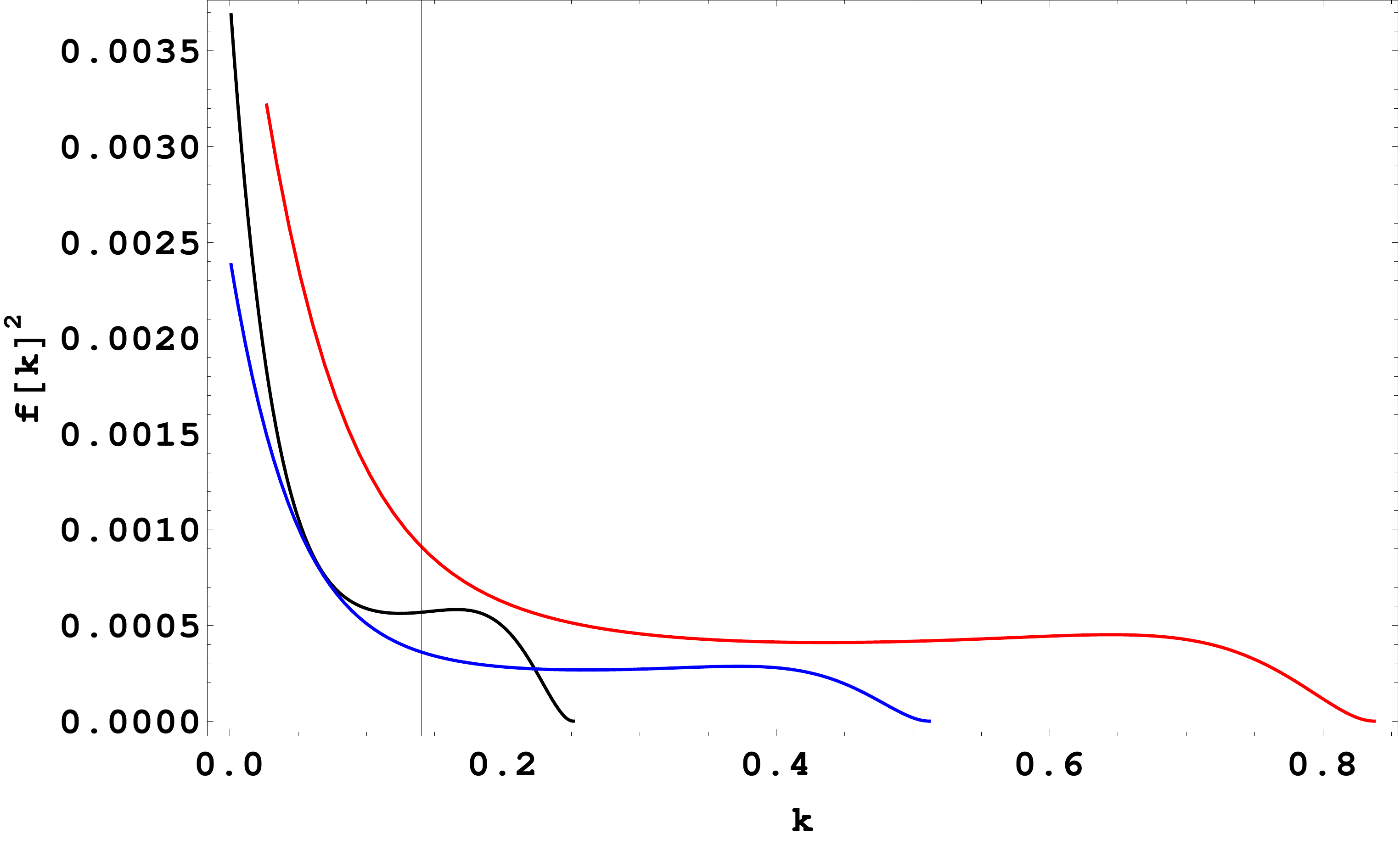}
  \caption{The symmetric power spectra $f^2(k)$ for {\color{red} $n_S=8/,n_A=9$}, {\color{black} $n_S=4/,n_A=5$} and {\color{blue} $n_S=6/,n_A=7$}}
  \label{simetri1}
\end{figure}

\begin{figure}[h!]
  \centering
    \includegraphics[width=.8\textwidth]{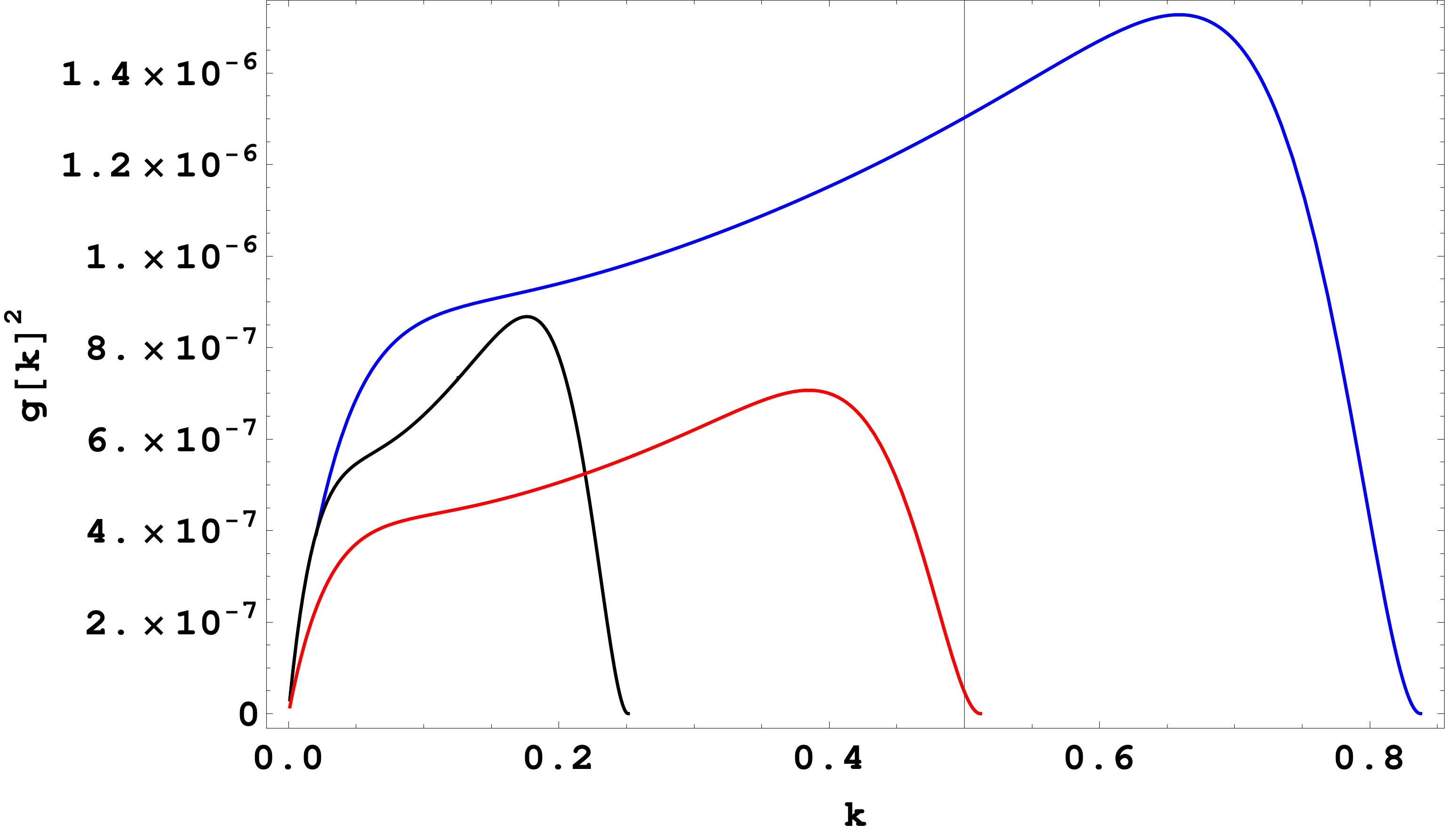}
  \caption{The helical power spectra $g^2(k)$ for {\color{red} $n_S=6/,n_A=7$}, {\color{black} $n_S=4/,n_A=5$} and {\color{blue} $n_S=8/,n_A=9$}}
  \label{simetri2}
\end{figure}
The anisotropic trace-free part   written in Fourier space is given by
\begin{equation}
\Pi_{ij}=\int \frac{d^3k^\prime}{2(2\pi)^4}\left[B_i(k^\prime)B_j(k-k^\prime)-\frac{\delta_{ij}}{3}B_l(k^\prime)B^l(k-k^\prime) \right],\nonumber
\end{equation}
and working with the  linear cosmological perturbation theory, one can find that anisotropic tensor is source of gravitational waves (GWs)  \cite{ca}
\begin{equation} \label{gw}
\ddot{h}+2H\dot{h}+k^2h\sim \Pi,
\end{equation}
where $h$ is the tensor part in the metric associated with GWs and $H$ the Hubble parameter.  Now, we define the  two point correlation function for the anisotropic trace-free part as
\begin{equation}\label{aniso}
\langle \Pi_{ij}^{(T)}(k)\Pi_{ij}^{(T)*}(k^\prime)\rangle=4(2\pi)^3 (f^2(k)+2ig^2(k))\delta^3(k-k^\prime),
\end{equation}
hence, using the Wick theorem and the later equations one can arrives to 
\begin{eqnarray}
g^2(k)&=&\frac{8}{512\pi^5}\int d^3k^\prime[\beta(1+\gamma^2)]S(k^\prime)A(\left|\textbf{k}-\textbf{k}^\prime\right|),\\
f^2(k)&=&\frac{1}{512\pi^5}\int d^3k^\prime[(1+2\gamma^2+\gamma^2\beta^2)S(k^\prime)S(\left|\textbf{k}-\textbf{k}^\prime\right|)
+4A(k^\prime)A(\left|\textbf{k}-\textbf{k}^\prime\right|)\gamma\beta],
\end{eqnarray}
with $
\beta=\frac{\textbf{k}\cdot(\textbf{k}-\textbf{k}^\prime)}{k\left|\textbf{k}-\textbf{k}^\prime\right|},  \quad  \gamma=\frac{\textbf{k}\cdot\textbf{k}^\prime}{kk^\prime}$, which is in agreement with \cite{ca}. The power spectra for symmetric and helical part are shown in figures \ref{simetri1} and \ref{simetri2}.

For studying the behavior of GW induced by  PMFs, we use the equation (\ref{gw}), obtaining the  figure \ref{correlators}.
\begin{figure}
  \centering
    \includegraphics[width=.8\textwidth]{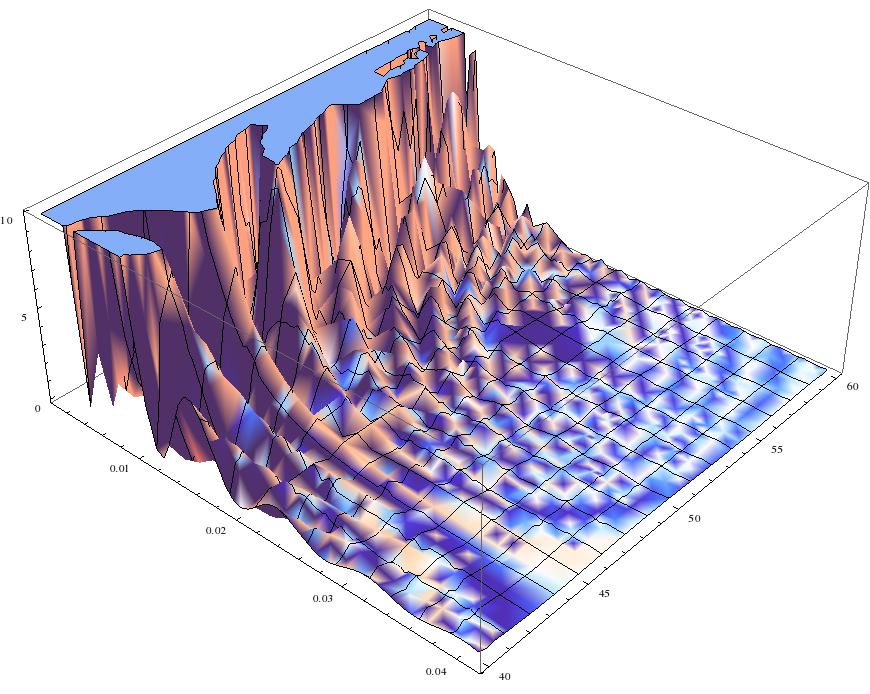}
  \caption{The gravity wave power spectra produced by PMFs.}
  \label{correlators}
\end{figure}
Here we can see the behavior of GW spectra in the interest scale. Basically, we have found that GWs spectra is constant in super-horizon scale  (type Harrison-Zeldovich spectrum) whilst at small scales, the spectra oscillates rapidly and  decay within time. 

\section{Discussion}
In this paper we have shown the exact solution of power spectra generated by helical primordial magnetic fields
created after inflation (causal) for some spectral indices. These fields could be originated by processes
in the early universe. More specifically, we studied  some mechanisms which generate magnetic fields via baryon asymmetry of the universe.
These fields are interesting in sense that creation of magnetic fields comes accompanied by helicity, therefore,
the dynamo mechanisms are more efficient and could drive to the explanation of large magnetic fields observed today.

\bibliographystyle{harvard}
\bibliography{Eduard}

\end{document}